\documentclass[11pt]{article}

\usepackage{acl}

\usepackage{times}
\usepackage{latexsym}

\usepackage[T1]{fontenc}
\usepackage[utf8]{inputenc}

\usepackage{microtype}
\usepackage{inconsolata}

\usepackage{graphicx}
\usepackage{amsmath}
\usepackage{amsfonts}
\usepackage{booktabs}
\usepackage{url}
\usepackage{xcolor}
\usepackage{nicefrac}
\usepackage{graphicx}
\usepackage{multirow}
\usepackage{float}
\usepackage{placeins}
\usepackage{cuted}
\usepackage{caption}
\usepackage{capt-of}

\title{\textsc{LLM-ALSO}: LLM-Driven Adaptive Learning-Signal Optimization for Multi-Agent Reinforcement Learning}

\author{
Xiaoguang Wu$^{1}$ \quad Zhi Zheng$^{1*}$ \quad Hui Xiong$^{2,3*}$ \\
$^{1}$University of Science and Technology of China \\
$^{2}$Thrust of Artificial Intelligence, The Hong Kong University of Science and Technology (Guangzhou) \\
$^{3}$Department of Computer Science and Engineering, The Hong Kong University of Science and \\
Technology, Hong Kong SAR, China \\
\texttt{wxg1500@mail.ustc.edu.cn} \quad
\texttt{zhengzhi97@ustc.edu.cn} \quad
\texttt{xionghui@ust.hk}
}

\begin{document}

\maketitle

\begin{abstract}
Effective training-time guidance is central to multi-agent reinforcement learning (MARL), yet remains difficult in sparse-reward settings where weak supervision limits coordination and policy improvement, and existing methods often require substantial domain expertise or manual design effort. Large language models (LLMs) provide a promising alternative for flexible learning-signal design, yet existing LLM-based methods remain largely single-agent-oriented, one-shot, or weakly validated for the evolving training dynamics of cooperative MARL.
To address these limitations, we propose \textsc{LLM-ALSO}, an iterative LLM-driven adaptive learning-signal optimization framework for MARL.
Rather than directly deploying LLM-generated rewards, \textsc{LLM-ALSO} decomposes adaptation into iterative diagnosis, proposal, and validation: a Critic LLM diagnoses stage-specific learning and coordination failures from sparse-return metrics and compact behavior evidence, a Generator LLM proposes candidate reward-shaping configurations conditioned on the diagnosis, and branch-validation feedback refines candidates before they affect the main training trajectory.
Through short-horizon validation and stage-aware adaptation, \textsc{LLM-ALSO} promotes only validated updates into training, reducing the risk of unreliable LLM-generated modifications.
Experiments on sparse-reward cooperative MARL tasks show that \textsc{LLM-ALSO} improves sparse-evaluation performance and learning efficiency.
The code is available at \url{https://github.com/xcGH-stu/adaptive-marl-signals}.
\end{abstract}

\section{Introduction}
Multi-agent reinforcement learning (MARL) provides a general framework for cooperative sequential decision-making and has been widely studied in domains such as robotic warehouses, traffic coordination, and multi-robot systems \citep{papoudakis2021benchmarking,chu2019traffic,krnjaic2024warehouse}. 
However, learning effective multi-agent policies remains challenging in sparse-reward settings, where delayed and limited supervision makes coordinated behaviors difficult to discover \citep{jeon2022maser,liu2023lazy}. 
Successful training in such settings therefore depends critically on effective training-time learning signals, especially informative reward-shaping signals that support exploration, coordination, and policy improvement \citep{wang2022irat}.

\begin{figure}[t]
    \centering
    \includegraphics[width=\linewidth]{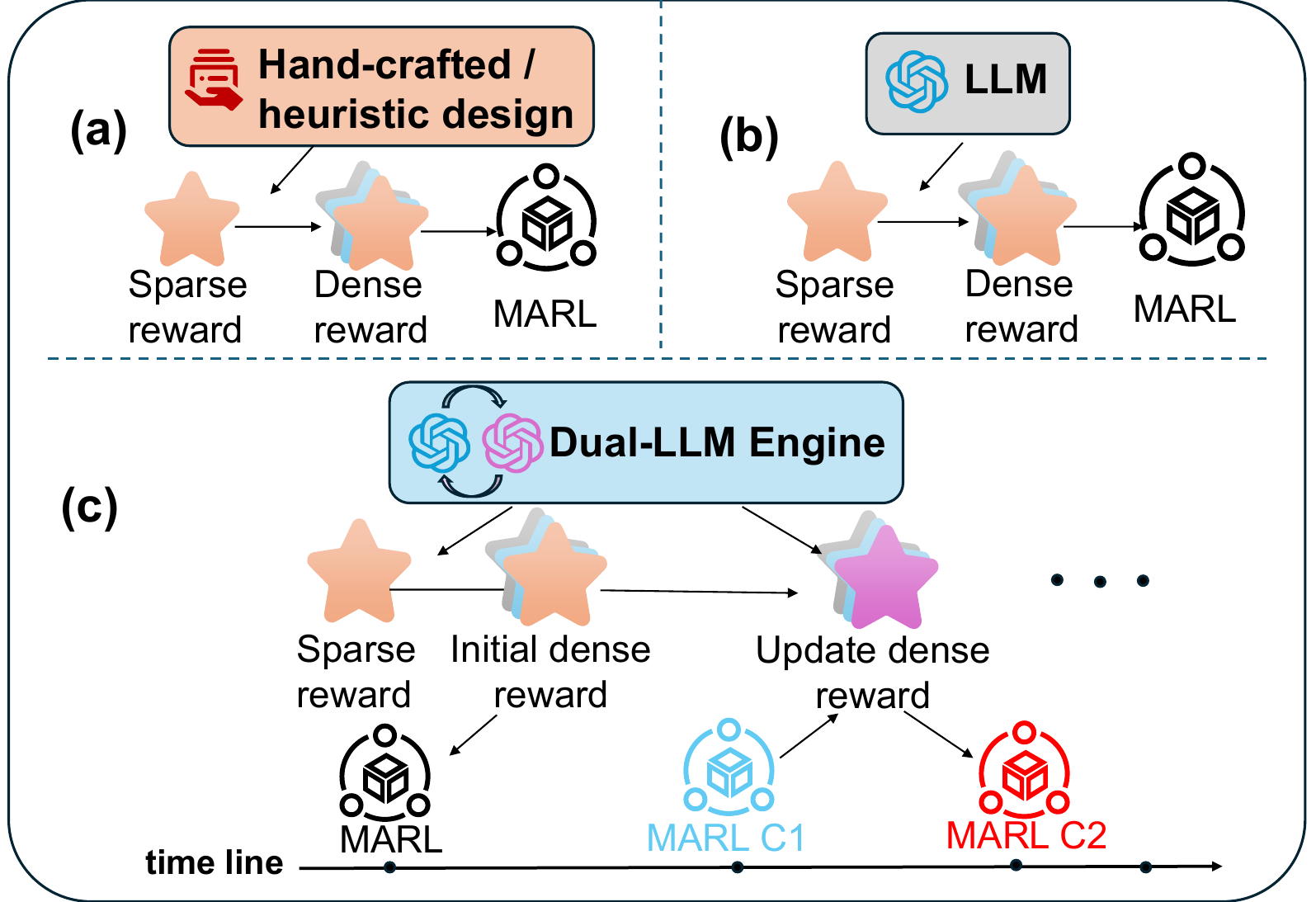}
    \caption{
    Comparison of reward-shaping paradigms in sparse-reward MARL.
    Compared to hand-crafted heuristic reward shaping (a) and single-stage LLM-based reward shaping (b), which keep the dense reward fixed after initialization, \textsc{LLM-ALSO} (c) enables stage-aware reward adaptation through initial reward selection and branch-validated adaptive updates.
    }
    \label{fig:intro_framework}
\end{figure}

Designing such learning signals for MARL is inherently difficult. 
Reward functions, in particular, serve as the primary source of supervision but often require substantial domain expertise, repeated trial-and-error, and environment-specific tuning. 
Prior work has explored hand-crafted reward shaping, auxiliary rewards, intrinsic motivation, reward estimation, and credit-assignment-aware objectives to provide denser supervision for cooperative agents \citep{du2019liir,mguni2022ligs,hu2022dre,liu2023lazy}. 
While effective in certain settings, these approaches typically rely on predefined signal forms, manually selected objectives, or algorithm-specific assumptions, making it difficult to adapt learning signals to the evolving coordination needs of MARL training.

Recent advances have explored LLMs as flexible sources of training guidance for reinforcement learning. 
In single-agent settings, LLMs have been used as proxy reward interfaces, executable reward-code generators, and iterative reward optimizers based on task descriptions, environment information, and feedback \citep{kwon2023reward,ma2024eureka,xie2024text2reward}.
In MARL, recent studies further use LLMs for reward generation, feedback interpretation, and coordination support during multi-agent training \citep{zhu2025lamarl,wang2025m3hf,li2025toolkit,li2024langground,godfrey2024marlin,siedler2025llm}. 
Despite these promising results, existing studies often treat reward design and policy guidance separately, relying on one-shot priors, predefined templates, or directly deployed LLM-generated signals. 
As a result, they rarely provide a closed-loop workflow that validates candidate learning-signal updates with training feedback and adapts them across stages of MARL training. 

Figure~\ref{fig:intro_framework} summarizes this gap: hand-crafted shaping and single-stage LLM reward generation typically keep dense signals fixed, rather than validating and adapting updates across training stages.
To address this limitation, we propose \textsc{LLM-ALSO}, an iterative LLM-driven adaptive learning-signal optimization framework for sparse-reward cooperative MARL.
Rather than directly deploying LLM-generated rewards, \textsc{LLM-ALSO} iterates over diagnosis, proposal, and validation.
At each stage, a Critic LLM identifies learning and coordination failures from training feedback and compact behavior evidence, while a Generator LLM proposes candidate reward-shaping configurations conditioned on the diagnosis.
Behavior evidence is used only as diagnostic context, helping the LLMs identify which reward-shaping directions should be explored without directly controlling agents.
Candidate reward updates are then evaluated through short-horizon branch validation from representative checkpoints before being promoted into the main training trajectory.
This design enables stage-aware reward adaptation while preserving the original sparse reward as the evaluation objective.

We evaluate \textsc{LLM-ALSO} on representative sparse-reward cooperative MARL tasks from Level-Based Foraging (LBF) \citep{papoudakis2021benchmarking}, covering multiple environment configurations with different coordination difficulties.
We instantiate the framework with representative MARL learners from different paradigms, including QMIX and MAPPO, and compare it against sparse-reward training, fixed dense-reward shaping, and single-LLM reward-generation baselines.
Experimental results demonstrate that adaptive learning-signal optimization improves sparse-evaluation performance and learning efficiency in LBF.

Overall, we make the following contributions:
\begin{itemize}
    \item We formulate learning-signal design in sparse-reward cooperative MARL as a stage-aware adaptive optimization problem, where reward shaping is refined according to training-stage-specific learning and coordination failures.

    \item We propose \textsc{LLM-ALSO}, an iterative Generator--Critic framework that uses behavioral diagnosis, candidate reward-shaping proposal, and short-horizon branch validation to adapt training signals while avoiding direct deployment of unvalidated LLM outputs.

    \item We evaluate \textsc{LLM-ALSO} on multiple LBF configurations under the original sparse task reward, showing improvements over sparse-reward training, fixed reward shaping, and single-LLM reward-generation baselines.
\end{itemize}
\section{Related Work}

Prior work related to our approach falls into three lines: non-LLM learning-signal design in MARL, LLM-based reward design for reinforcement learning, and LLM-guided MARL coordination.

\textbf{Non-LLM learning signals in MARL.}
MARL learning-signal design has long relied on auxiliary rewards, intrinsic motivation, reward estimation, and credit-assignment-aware objectives.
Representative methods include LIIR for learning individual intrinsic rewards under team rewards \citep{du2019liir}, LIGS for learnable intrinsic-reward generation in sparse coordination tasks \citep{mguni2022ligs}, DRE-MARL for distributional reward estimation \citep{hu2022dre}, and LAIES for diligence-based intrinsic motivation \citep{liu2023lazy}.
These methods show that auxiliary signals can improve exploration and credit assignment, but their signal forms are usually predefined or tied to specific modeling assumptions, rather than adaptively inferred from training feedback and stage-specific coordination failures.

\textbf{LLM-based reward design for reinforcement learning.}
Recent work uses LLMs or vision-language models to reduce manual reward engineering.
Reward Design with Language Models, Text2Reward, and Eureka generate or optimize rewards from task descriptions, environment representations, or iterative feedback \citep{kwon2023reward,xie2024text2reward,ma2024eureka}, while VLM-RM uses pretrained vision-language models as zero-shot reward models \citep{rocamonde2024vlm}.
However, these methods mainly target single-agent robotic or embodied-control settings, and do not directly address cooperative MARL challenges such as credit assignment, non-stationarity, and stage-dependent coordination issues.

\textbf{LLM-guided MARL coordination and adaptation.}
Recent studies have begun to combine LLMs with MARL, including language-grounded communication for ad-hoc teamwork \citep{li2024langground}, LLM-generated policy and reward priors for cooperative multi-robot learning \citep{zhu2025lamarl}, LLM-parsed multi-phase human feedback for reward updates \citep{wang2025m3hf}, and LLM-mediated training interventions \citep{siedler2025llm}.
These methods typically use LLMs as communication grounders, prior generators, feedback parsers, or external intervention modules.
In contrast, our framework uses LLMs to propose structured reward-shaping candidates from training and behavior evidence, validates them through short-horizon branches against a no-change control, and adapts learning signals across training stages.

\section{Method}
\label{sec:method}

\subsection{Problem Setup and Method Overview}
\label{sec:method_overview}

We consider a cooperative multi-agent reinforcement learning (MARL) problem with $N$ agents, formulated as a Dec-POMDP
$\mathcal{M}=\langle \mathcal{S}, \{\mathcal{A}_i\}_{i=1}^{N}, \{\mathcal{O}_i\}_{i=1}^{N}, P, r^{env}, \gamma \rangle$.
At timestep $t$, each agent $i$ receives a local observation $o_t^i$, selects an action $a_t^i$, and the joint action
$\mathbf{a}_t=(a_t^1,\ldots,a_t^N)$ induces a transition and a shared sparse task reward $r_t^{env}$.
The final objective is always evaluated under the original environment reward:
\begin{equation}
J^{env}(\pi)=
\mathbb{E}_{\pi}\left[
\sum_{t=0}^{T}\gamma^t r_t^{env}
\right].
\label{eq:env_objective}
\end{equation}

Sparse cooperative tasks often provide insufficient intermediate feedback for exploration, role allocation, and coordinated task completion.
We therefore allow reward shaping during training while keeping the evaluation objective unchanged.
During training, the learner receives
\begin{equation}
r_t^{train}=r_t^{env}+r_t^{shape},
\label{eq:train_reward}
\end{equation}
where $r_t^{shape}$ is used only for policy learning and is disabled during evaluation.
All reported performance metrics are computed using $r_t^{env}$, so shaping can affect learning but not the evaluation criterion.

\textsc{LLM-ALSO} uses LLMs in a constrained way.
The LLMs do not directly control agents, modify the learner, or select actions.
Instead, they serve as evidence-conditioned diagnosis and proposal modules over a structured potential-based reward-shaping space.
A proposed shaping configuration can influence the main trajectory only after it satisfies the structured PBRS parameterization and is empirically validated by short-horizon branches.

\begin{figure*}[t]
    \centering
    \includegraphics[width=\textwidth]{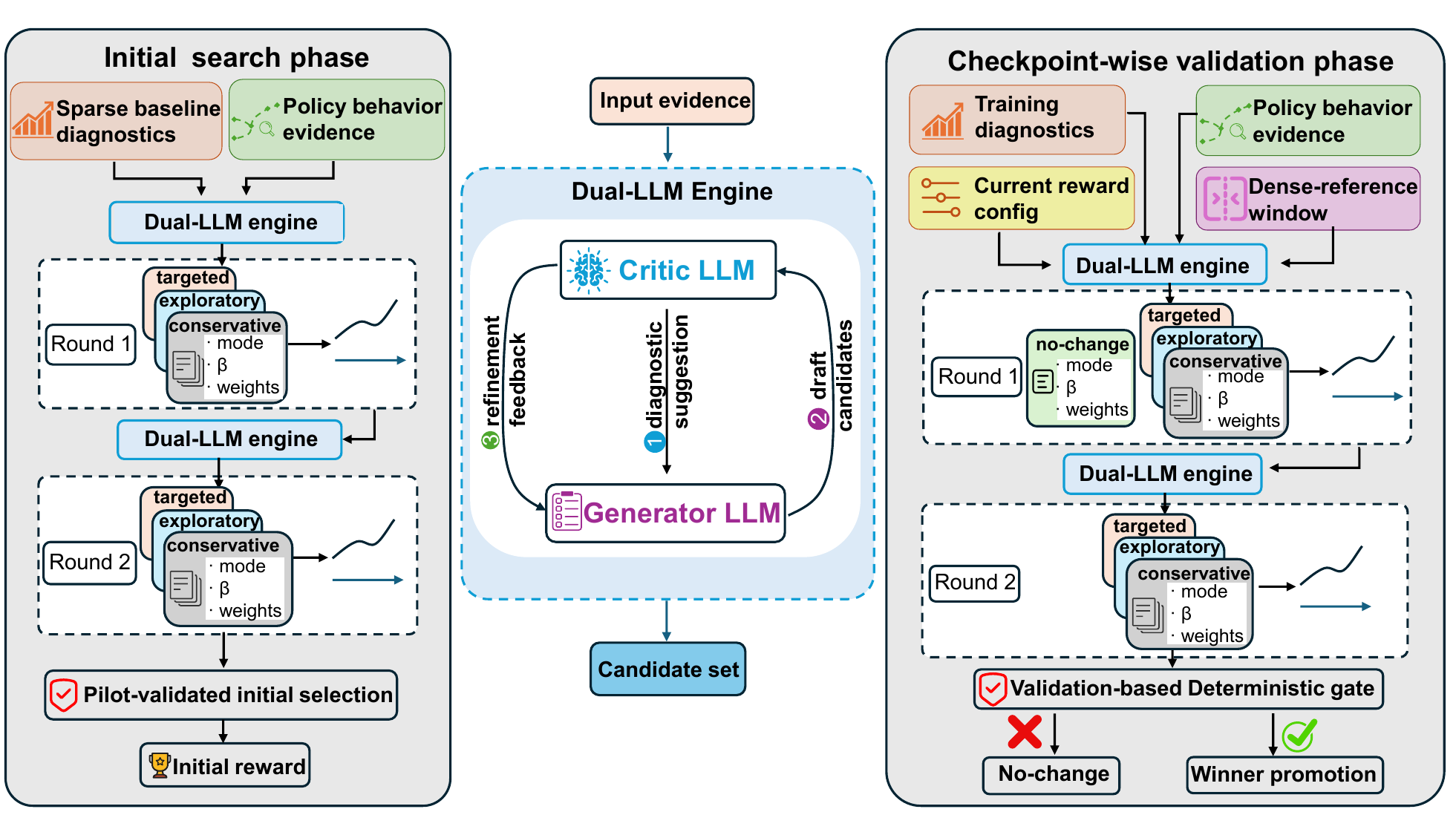}
    \caption{
    Overview of \textsc{LLM-ALSO}.
    A shared dual-LLM engine generates structured PBRS candidates from training diagnostics and behavior evidence.
    The initial search phase selects an early shaping configuration, while checkpoint-wise validation tests later updates against a no-change control before promoting validated branches into the mainline.
    }
    \label{fig:method_overview}
\end{figure*}

Figure~\ref{fig:method_overview} provides an overview of this two-phase workflow and the shared dual-LLM engine.
The workflow has two phases.
First, the initial search phase uses sparse diagnostics and behavior evidence to select an early reward-shaping configuration and endpoint checkpoint that can bootstrap sparse-reward learning.
This selected pair initializes both a fixed-reference continuation and the adaptive mainline.
Second, the checkpoint-wise validation phase revisits the shaping configuration at a small number of intervention checkpoints.
LLM-proposed updates are tested through short-horizon branches against a mandatory no-change control, and a deterministic validation gate decides whether to retain the current configuration or promote a winning update branch.
Thus, \textsc{LLM-ALSO} separates proposal from adoption: LLMs generate structured candidates, while empirical branch outcomes determine whether any candidate is used.

\subsection{Structured PBRS Configuration Space}
\label{sec:pbrs_search_space}

We instantiate the training-time shaping signal using potential-based reward shaping (PBRS)~\citep{ng1999policy}:
\begin{equation}
r_t^{shape}
=
\beta
\left(
\gamma \Phi_{\theta}(\tilde{s}_{t+1})
-
\Phi_{\theta}(\tilde{s}_{t})
\right),
\label{eq:pbrs}
\end{equation}
where $\beta$ controls the shaping strength and
$\tilde{s}_t=(s_t,h_t)$ is an augmented training-time state.
The auxiliary cache $h_t$ stores lightweight episode-level information, such as visited regions, recent failed collection attempts, target assignments, or other task-specific behavioral events.
It is used only for computing the shaping potential and is not part of the evaluation objective.
Because $\tilde{s}_t$ may include training-time cache variables, we treat this formulation as a constrained and interpretable PBRS-style shaping family, without assuming the standard policy-invariance guarantee under arbitrary state augmentation.

Rather than allowing arbitrary dense reward code, we restrict candidate generation to a structured PBRS configuration space.
Each potential function is represented as a weighted combination of interpretable task components:
\begin{equation}
\Phi_{\theta}(\tilde{s})
=
\sum_{k\in\mathcal{K}} w_k \phi_k(\tilde{s}),
\label{eq:pbrs_potential}
\end{equation}
where $w_k \ge 0$ for all $k\in\mathcal{K}$ and $\sum_{k\in\mathcal{K}} w_k=1$.
Here $\mathcal{K}$ is instantiated for Level-Based Foraging with components for collection progress, approach behavior, coverage and discovery, target allocation, joint collection readiness, and late-stage stability.
Each basis function $\phi_k$ captures a property of sparse cooperative foraging.
This parameterization keeps the search space interpretable and reduces unconstrained reward hacking.

A shaping configuration is denoted by
\begin{equation}
\theta=(\beta,m,\mathbf{w}),
\label{eq:pbrs_config}
\end{equation}
where $m$ is a high-level shaping mode and $\mathbf{w}$ gives the component weights.
The mode constrains the candidate toward a stage-appropriate direction, such as early discovery, collection readiness, allocation balance, or stability, while the weights determine each component's contribution to the potential.
All Generator proposals must remain inside this configuration space.

\subsection{Behavior Evidence and Dual-LLM Roles}
\label{sec:behavior_llm_roles}

The structured PBRS space defines valid candidates.
We next describe how evidence is converted into candidate proposals.
\textsc{LLM-ALSO} uses a dual-LLM interface with a Critic and a Generator.
The Critic diagnoses learning failures from reward metrics and compact behavior evidence, while the Generator proposes valid PBRS configurations within the search space in Section~\ref{sec:pbrs_search_space}.

Behavior evidence is extracted from evaluation episodes and used only as diagnostic context, without directly controlling actions, modifying the learner, or changing the environment dynamics.
It summarizes coverage, near-target movement, failed collection attempts, target concentration, allocation imbalance, and stability indicators.
Rather than passing raw trajectories, the framework aggregates these signals into compact evidence keys.

The Critic combines behavior summaries with recent training and validation metrics and the current shaping configuration.
Its output is an integrated guidance card that specifies diagnosed failure modes, supporting evidence keys, candidate-generation constraints, and potential risks.
Conditioned on this card, the Generator proposes a small set of structured PBRS candidates.
To cover the local search space, these candidates are organized into complementary proposal roles, such as targeted updates that address the diagnosed bottleneck, exploratory alternatives that test a different shaping direction, and conservative adjustments that stay close to the current configuration.
The Generator is restricted to choosing a shaping mode, a strength parameter, and component weights; it cannot introduce arbitrary reward terms or policy-level interventions.

This separation of roles is central to the framework.
The Critic turns performance and behavior evidence into an interpretable diagnosis, and the Generator converts this diagnosis into valid reward-shaping candidates.
Whether a candidate is adopted is deferred to empirical validation in the two phases described below.

\subsection{Initial Reward-Shaping Search}
\label{sec:initial_search}

The initial search phase selects an early reward-shaping configuration that can bootstrap exploration and coordination under sparse rewards, rather than a globally optimal dense reward for the entire training process.
It also selects an endpoint checkpoint from which later training can continue.

The phase starts from a sparse diagnostic run.
From this run, the framework collects sparse-reward learning curves and compact behavior evidence.
The Critic diagnoses early learning bottlenecks and links them to the structured PBRS components, and the Generator proposes candidate configurations specified by a shaping strength, a mode, and component weights.

We use a two-round search procedure.
In the first round, the Generator proposes diverse candidate directions, each of which is trained under a limited pilot budget to produce a short trajectory and an endpoint checkpoint.
The Critic then reviews the pilot outcomes using sparse-return metrics and behavior changes.
The second round refines the search by exploiting promising directions and correcting observed weaknesses.

After both rounds, the framework selects an initial shaping configuration and endpoint checkpoint, denoted by $(\theta_0,c_0)$.
The selection considers sparse task performance, learning stability, and consistency with the diagnosis.
Both the fixed reference and adaptive mainline start from the same selected checkpoint $c_0$ and shaping configuration $\theta_0$.
This shared source separates the initial search benefit from later checkpoint-wise replacement.

\subsection{Checkpoint-wise Validation and Replacement}
\label{sec:checkpoint_validation}

The checkpoint-wise validation phase revisits the shaping signal at fixed intervention checkpoints.
Its goal is not to accept every LLM-proposed update, but to test whether a local reward-shaping change improves over continuing with the existing configuration.
Thus, each checkpoint compares update branches against a local no-change control.

Let $C_j$ denote the $j$-th intervention checkpoint, where the current mainline checkpoint and shaping configuration are $(c_j,\theta_j)$.
The framework builds a checkpoint context from recent sparse-return metrics, compact behavior evidence, branch history, the current shaping configuration, and, when available, a fixed-reference window.
Given this context, the Critic diagnoses the current training stage, and the Generator proposes a small set of structured PBRS update candidates.

Each update branch starts from $c_j$ and is trained for a short horizon.
The no-change control continues from the same checkpoint without modifying the shaping configuration, providing the local counterfactual for comparison.
Branch outcomes from the first round are returned to the dual-LLM interface, allowing the second round to refine candidate generation based on observed evidence.

The final replacement decision is made by a deterministic validation gate, not by the LLM.
The gate uses sparse-return and stability criteria, promoting an update only when it improves over the no-change control without unstable or transient gains.
If no update passes, the current shaping configuration is retained.

The next mainline state is obtained by winner-branch promotion.
If the no-change control wins, training continues with the same shaping configuration; if an update branch wins, its endpoint checkpoint and shaping configuration are promoted into the mainline.
Thus, LLMs propose structured PBRS updates, while empirical branch validation determines whether any update is adopted.

\section{Experiments}
\label{sec:experiments}

We evaluate \textsc{LLM-ALSO} on sparse-reward cooperative MARL tasks from Level-Based Foraging (LBF).
Our experiments examine whether \textsc{LLM-ALSO} improves sparse-evaluation performance over sparse training, fixed reward shaping, and single-LLM reward generation across value-based and policy-gradient learners.
We further analyze how the staged workflow, including initial reward-shaping selection and checkpoint-wise branch validation, contributes to the final training trajectory.

\subsection{Experimental Setup}
\label{sec:exp_setup}

\paragraph{Benchmarks.}
We evaluate \textsc{LLM-ALSO} on cooperative sparse-reward tasks from Level-Based Foraging (LBF) \citep{papoudakis2021benchmarking}.
We use five LBF configurations with different map sizes, numbers of agents, and numbers of food items: 8$\times$8-2p-1f, 8$\times$8-2p-2f, 2s-8$\times$8-2p-2f, 10$\times$10-2p-1f, and 15$\times$15-3p-4f.
These tasks require agents to explore, approach feasible food, allocate agents to targets, and coordinate joint collection under sparse task feedback.
Throughout all experiments, evaluation is performed using the original sparse environment reward; auxiliary dense rewards are used only as training-time learning signals and are disabled during evaluation.

\paragraph{Algorithms.}
We instantiate \textsc{LLM-ALSO} with both value-based and policy-gradient MARL algorithms.
Specifically, we use QMIX~\citep{rashid2018qmix} as a representative value-based method and MAPPO~\citep{yu2022mappo} as a representative policy-gradient method.
This allows us to evaluate whether the proposed learning-signal optimization workflow applies across different MARL optimization paradigms within the same sparse cooperative benchmark family.
Within each comparison group, all methods use the same MARL algorithm and differ only in the learning signal.

\paragraph{Compared methods.}
We compare \textsc{LLM-ALSO} with three baselines: Sparse, which uses only the original environment reward; Fixed-RS, which uses a predefined structured PBRS signal; and Single-LLM-RG, which selects one PBRS configuration using a single LLM and then keeps it fixed.
All dense-signal methods use the same structured PBRS configuration space, while \textsc{LLM-ALSO} further updates the configuration through dual-LLM branch validation.
Within each learner--environment setting, all methods share the same MARL learner, final training horizon, and sparse-evaluation protocol; detailed workflow accounting is provided in Appendix~\ref{app:workflow_budgets}.

\paragraph{Evaluation metrics.}
All metrics are computed using the original sparse environment reward.
We report maximum test sparse return and average test sparse return, measuring peak task performance and overall learning efficiency, respectively.
For multi-seed experiments, we report mean and standard deviation across random seeds.
Trajectory construction details for staged workflows are provided in Appendix~\ref{app:workflow_budgets}.

\subsection{Main Results}
\label{sec:main_results}

Table~\ref{tab:main_results} reports the main sparse-evaluation results across LBF tasks and MARL learners.
Using the maximum and average test sparse return defined above, we compare \textsc{LLM-ALSO} with Sparse, Fixed-RS, and Single-LLM-RG to evaluate both peak performance and overall learning efficiency.
For compactness, Single-LLM-RG is abbreviated as S-LLM-RG in Table~\ref{tab:main_results}.

\begin{table}[H]
\centering
\scriptsize
\setlength{\tabcolsep}{2.0pt}
\renewcommand{\arraystretch}{1.02}
\resizebox{\linewidth}{!}{
\begin{tabular}{@{}llcccc@{}}
\toprule
\multirow{2}{*}{\textbf{Config.}} &
\multirow{2}{*}{\textbf{Method}} &
\multicolumn{2}{c}{\textbf{QMIX}} &
\multicolumn{2}{c}{\textbf{MAPPO}} \\
\cmidrule(lr){3-4}\cmidrule(lr){5-6}
& & \textbf{Max} & \textbf{Avg.} & \textbf{Max} & \textbf{Avg.} \\
\midrule

\multirow{4}{*}{\shortstack[l]{8$\times$8\\2p-1f}}
& Sparse        & $1.00{\pm}0.00$ & $0.47{\pm}0.02$ & $1.00{\pm}0.00$ & $0.89{\pm}0.02$ \\
& Fixed-RS      & $1.00{\pm}0.01$ & $0.47{\pm}0.07$ & $1.00{\pm}0.00$ & $0.90{\pm}0.02$ \\
& S-LLM-RG      & $1.00{\pm}0.00$ & $0.53{\pm}0.06$ & $1.00{\pm}0.00$ & $0.89{\pm}0.02$ \\
& \textsc{LLM-ALSO} 
                & $1.00{\pm}0.00$ & $\mathbf{0.83{\pm}0.05}$ 
                & $1.00{\pm}0.00$ & $\mathbf{0.92{\pm}0.03}$ \\
\midrule

\multirow{4}{*}{\shortstack[l]{8$\times$8\\2p-2f}}
& Sparse        & $1.00{\pm}0.01$ & $0.53{\pm}0.02$ & $1.00{\pm}0.00$ & $0.86{\pm}0.05$ \\
& Fixed-RS      & $1.00{\pm}0.00$ & $0.52{\pm}0.03$ & $1.00{\pm}0.00$ & $\mathbf{0.87{\pm}0.02}$ \\
& S-LLM-RG      & $1.00{\pm}0.00$ & $0.55{\pm}0.06$ & $1.00{\pm}0.00$ & $0.85{\pm}0.02$ \\
& \textsc{LLM-ALSO} 
                & $1.00{\pm}0.00$ & $\mathbf{0.61{\pm}0.08}$ 
                & $1.00{\pm}0.00$ & $0.86{\pm}0.02$ \\
\midrule

\multirow{4}{*}{\shortstack[l]{2s-8$\times$8\\2p-2f}}
& Sparse        & $1.00{\pm}0.00$ & $0.68{\pm}0.05$ & $1.00{\pm}0.00$ & $0.86{\pm}0.05$ \\
& Fixed-RS      & $1.00{\pm}0.00$ & $0.73{\pm}0.04$ & $1.00{\pm}0.00$ & $0.87{\pm}0.02$ \\
& S-LLM-RG      & $1.00{\pm}0.00$ & $0.75{\pm}0.08$ & $1.00{\pm}0.00$ & $0.87{\pm}0.04$ \\
& \textsc{LLM-ALSO} 
                & $1.00{\pm}0.00$ & $\mathbf{0.76{\pm}0.02}$ 
                & $1.00{\pm}0.00$ & $\mathbf{0.90{\pm}0.02}$ \\
\midrule

\multirow{4}{*}{\shortstack[l]{10$\times$10\\2p-1f}}
& Sparse        & $0.85{\pm}0.01$ & $0.23{\pm}0.01$ & $1.00{\pm}0.00$ & $0.84{\pm}0.03$ \\
& Fixed-RS      & $0.76{\pm}0.09$ & $0.20{\pm}0.02$ & $1.00{\pm}0.00$ & $0.91{\pm}0.00$ \\
& S-LLM-RG      & $0.83{\pm}0.11$ & $0.25{\pm}0.05$ & $1.00{\pm}0.00$ & $0.93{\pm}0.03$ \\
& \textsc{LLM-ALSO} 
                & $\mathbf{0.95{\pm}0.03}$ & $\mathbf{0.32{\pm}0.00}$ 
                & $1.00{\pm}0.00$ & $\mathbf{0.96{\pm}0.00}$ \\
\midrule

\multirow{4}{*}{\shortstack[l]{15$\times$15\\3p-4f}}
& Sparse        & $0.05{\pm}0.01$ & $0.03{\pm}0.01$ & $0.75{\pm}0.12$ & $0.42{\pm}0.10$ \\
& Fixed-RS      & $0.06{\pm}0.00$ & $0.03{\pm}0.00$ & $0.84{\pm}0.10$ & $0.55{\pm}0.05$ \\
& S-LLM-RG      & $0.05{\pm}0.01$ & $0.02{\pm}0.01$ & $0.85{\pm}0.03$ & $0.58{\pm}0.03$ \\
& \textsc{LLM-ALSO} 
                & $\mathbf{0.10{\pm}0.01}$ & $\mathbf{0.04{\pm}0.02}$ 
                & $\mathbf{0.98{\pm}0.03}$ & $\mathbf{0.72{\pm}0.04}$ \\
\bottomrule
\end{tabular}
}

\caption{
Main sparse-evaluation results on LBF.
Max and Avg. denote maximum and average test sparse return, reported as mean $\pm$ standard deviation over random seeds.
S-LLM-RG denotes Single-LLM-RG.
}
\label{tab:main_results}
\end{table}

Overall, \textsc{LLM-ALSO} achieves the highest average sparse return in most reported LBF--learner settings and remains competitive in the remaining easy setting where peak performance is already saturated.
Because several easier configurations reach near-perfect maximum return across methods, average sparse return is more informative for comparing learning speed and stability.
The gains are more pronounced for QMIX, where sparse-reward coordination is more challenging, while MAPPO shows smaller differences on saturated tasks but improvements on the harder 15$\times$15-3p-4f task.

\begin{figure}[H]
    \centering
    \includegraphics[width=\linewidth]{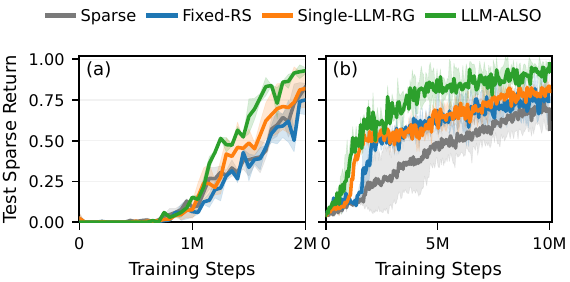}
    \caption{
    Representative sparse-evaluation learning curves on LBF:
    (a) QMIX 10$\times$10-2p-1f and
    (b) MAPPO 15$\times$15-3p-4f.
    Shading denotes standard deviation over seeds.
    }
    \label{fig:main_result}
\end{figure}

Figure~\ref{fig:main_result} complements the aggregate results by showing representative training dynamics.
In QMIX 10$\times$10-2p-1f, the methods are close early in training, but \textsc{LLM-ALSO} separates from the baselines in later stages and reaches a higher sparse return.
In MAPPO 15$\times$15-3p-4f, dense-signal methods improve over sparse training, while \textsc{LLM-ALSO} rises earlier and maintains the strongest sparse-evaluation performance for most of training.
Since all curves use sparse evaluation, the gains reflect task performance, not shaping rewards.

\subsection{Analysis of Staged Optimization}
\label{sec:staged_analysis}

We next analyze how the staged components of \textsc{LLM-ALSO} shape the final training trajectory.
The initial reward-shaping search provides an early dense signal, while checkpoint-wise validation decides whether later updates should replace or preserve the current configuration.

The main results suggest that the initial search can provide a useful foundation for sparse-reward learning.
For example, in the MAPPO 15$\times$15 setting in Figure~\ref{fig:main_result}, \textsc{LLM-ALSO} improves early learning, indicating that the selected shaping configuration helps sparse exploration.
However, the initial configuration is not always sufficient.
In QMIX 10$\times$10-2p-1f, \textsc{LLM-ALSO} does not obtain a clearly superior early trajectory from the initial configuration alone; its later improvement is consistent with stage-specific checkpoint adaptation.
To examine this process, we visualize both the short-horizon branch trajectories and the resulting validation margins at the two checkpoints.

\begin{figure}[H]
    \centering
    \includegraphics[width=\linewidth]{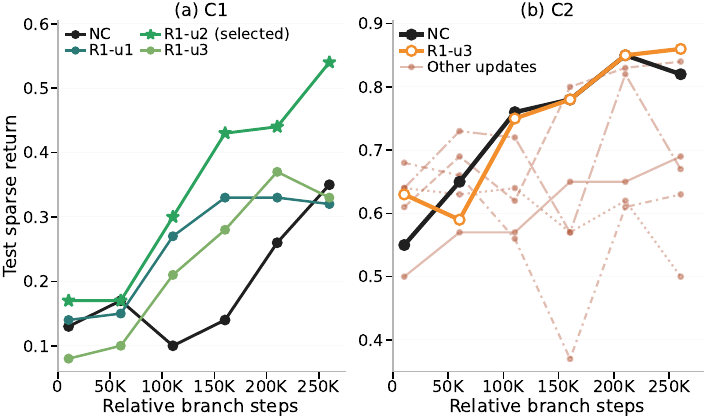}
   \caption{
Short-horizon branch validation curves for the representative QMIX 10$\times$10-2p-1f run.
C1 promotes R1-u2, whereas C2 retains no-change despite the near-tie R1-u3 update.
}
    \label{fig:branch_validation_curves}
\end{figure}

\begin{figure}[H]
    \centering
    \includegraphics[width=\linewidth]{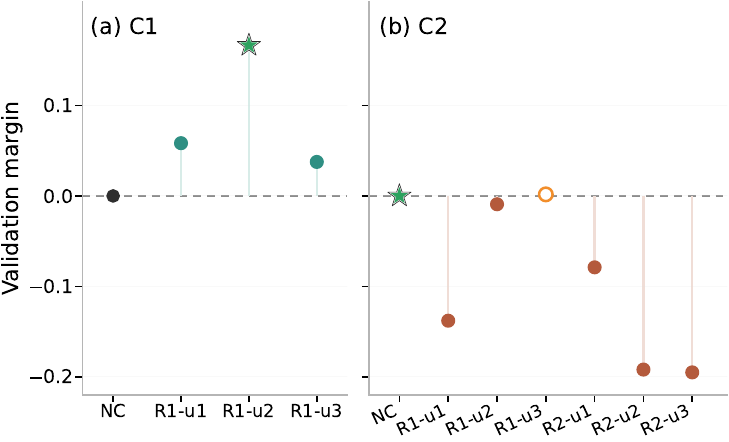}
   \caption{
Validation-margin summary for the representative QMIX 10$\times$10-2p-1f run.
Margins are relative to the local no-change control.
C1 promotes the starred branch; C2 retains no-change.
}
    \label{fig:checkpoint_validation}
\end{figure}

Figures~\ref{fig:branch_validation_curves} and~\ref{fig:checkpoint_validation} provide complementary views of checkpoint-wise validation.
The branch curves show the short-horizon behavior of each candidate, while the margin plot summarizes the same evidence relative to the local no-change control.
At C1, R1-u2 improves over the no-change control and obtains a clear positive margin, providing evidence for promoting the update into the mainline.
This is the positive case of stage-aware adaptation: the workflow does not merely continue the initial shaping configuration, but identifies a new configuration whose short-horizon behavior supports replacement at the current training stage.
At C2, the strongest update candidate remains close to no-change and the remaining candidates underperform, so the validation gate retains the current configuration.

Together, these results show how short-horizon validation turns LLM-proposed candidates into evidence-based adaptation decisions.
At C1, the workflow is able to identify a stage-appropriate update whose branch behavior supports replacement, showing that adaptive reward shaping can improve the subsequent training trajectory when the current configuration is no longer the best local choice.
At C2, the same validation mechanism prevents a weak or near-tied update from being adopted, even when it appears competitive with the no-change control.
Thus, the no-change branch is not only a safety fallback, but also the local counterfactual that makes the replacement decision meaningful.
This supports the central design of \textsc{LLM-ALSO}: LLM proposals are treated as hypotheses to be tested, and only candidates with sufficient short-horizon evidence are promoted into the mainline.

\subsection{Ablation on Dual-LLM Interaction}
\label{sec:dual_llm_ablation}

Having analyzed how the validation gate selects or rejects updates, we next examine the role of the dual-LLM engine that proposes those updates.
We compare \textsc{LLM-ALSO} with Single-LLM Adapt, a single-LLM adaptive variant, on QMIX 10$\times$10-2p-1f.
This setting is a representative ablation case because the initial shaping configuration alone does not yield a clearly superior early trajectory, making later adaptive updates important.
Single-LLM Adapt keeps the same staged adaptation workflow as \textsc{LLM-ALSO}, including the structured PBRS search space, short-horizon branch validation, and mainline promotion, but replaces the dual-LLM Critic--Generator interaction with a single LLM.
Thus, this comparison targets the role of the dual-LLM interaction within the adaptive workflow, rather than the one-shot Single-LLM-RG baseline.

\begin{figure}[H]
    \centering
    \includegraphics[width=\linewidth]{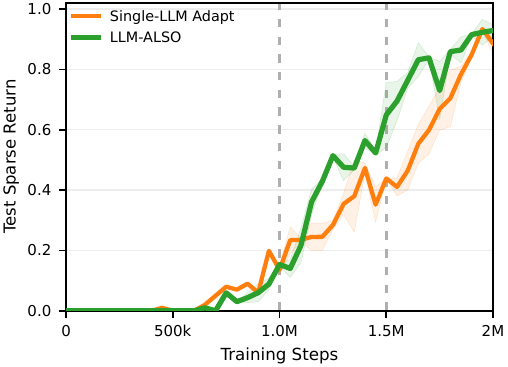}
    \caption{
    Ablation on dual-LLM interaction for QMIX 10$\times$10-2p-1f.
Curves show the mean, and shading denotes standard deviation.
    }
    \label{fig:dual_llm_ablation}
\end{figure}

\begin{table}[H]
\centering
\small
\setlength{\tabcolsep}{5pt}
\begin{tabular}{lcc}
\toprule
Method & Avg. & Max \\
\midrule
Single-LLM Adapt & $0.26{\pm}0.01$ & $0.94{\pm}0.01$ \\
\textsc{LLM-ALSO} & $\mathbf{0.32{\pm}0.00}$ & $\mathbf{0.95{\pm}0.03}$ \\
\bottomrule
\end{tabular}
\caption{
Dual-LLM interaction ablation on QMIX 10$\times$10-2p-1f.
Values report mean $\pm$ standard deviation.
}
\label{tab:dual_llm_ablation}
\end{table}

As shown in Figure~\ref{fig:dual_llm_ablation} and Table~\ref{tab:dual_llm_ablation}, the key difference between the two adaptive variants appears around the mid-training stage.
Together with the C1 validation evidence above, this post-checkpoint divergence suggests that the Critic--Generator interaction helps produce a stage-appropriate update candidate that passes the same validation gate and improves the subsequent mainline trajectory.
This pattern is consistent with the dual-LLM role separation in our method: the Critic turns sparse-return metrics and behavior evidence into a stage-specific diagnosis, while the Generator uses this diagnosis to propose constrained PBRS candidates.
Under the same staged workflow and validation mechanism, Single-LLM Adapt does not produce the same post-checkpoint improvement in this representative case.
Thus, the dual-LLM workflow mainly provides a stronger post-validation trajectory through more effective candidate generation, rather than a higher maximum return.

\section{Conclusion}

We presented \textsc{LLM-ALSO}, a stage-aware framework for optimizing training-time learning signals in sparse-reward cooperative MARL.
Rather than directly deploying LLM-generated rewards, \textsc{LLM-ALSO} uses a constrained dual-LLM engine to diagnose stage-specific learning difficulties and propose structured reward-shaping updates.
These updates affect the main trajectory only after short-horizon branch validation, separating LLM proposal from empirical adoption.
Experiments on LBF show that \textsc{LLM-ALSO} improves sparse-evaluation performance over sparse training, fixed reward shaping, and single-LLM reward generation in most settings, especially on harder coordination tasks.
Further analysis shows that checkpoint-wise validation supports effective replacement and safe retention, while a representative ablation suggests that dual-LLM interaction strengthens stage-aware candidate generation.
Overall, \textsc{LLM-ALSO} suggests a practical way to use LLMs for MARL reward shaping: constrain their proposals, adapt them across training stages, and validate them before adoption.

\clearpage
\section*{Limitations}

This study has several limitations.
Our experiments focus on LBF, so broader MARL benchmarks are needed to further test generality.
\textsc{LLM-ALSO} also relies on a predefined PBRS-style component space, which improves interpretability but makes performance depend on the expressiveness of these components.
In addition, checkpoint-wise validation requires extra short-horizon branch training, and local branch outcomes cannot perfectly predict long-term training behavior.
Future work can extend the framework to broader domains, automatically construct shaping components, and develop more efficient uncertainty-aware validation gates.

\section*{Ethical Considerations}

This work studies LLM-assisted learning-signal optimization for cooperative multi-agent reinforcement learning in simulated benchmark environments. The experiments do not involve human subjects, private user data, personally identifiable information, or real-world deployment. The LLM inputs used in our workflow consist of task descriptions, structured reward-shaping schemas, and aggregate training evidence, rather than sensitive natural-language data. The main ethical risk is that automatically generated reward-shaping signals may encode misspecified objectives or induce unintended behaviors if transferred to real multi-agent systems without sufficient validation. To mitigate this risk, our method constrains LLM outputs to a structured PBRS search space, validates candidate learning signals through short-horizon branch experiments with a no-change control, and evaluates performance only under the original sparse task reward. Our results should therefore be interpreted as evidence in controlled simulation settings, rather than as a guarantee of safe deployment in physical, human-facing, or safety-critical environments.
\bibliography{references}

\clearpage
\appendix

\section{Workflow Budgets and Trajectory Accounting}
\label{app:workflow_budgets}

\paragraph{Budget accounting.}
We use fixed workflow budgets for each learner--environment setting.
Most experiments use a $2.05$M-step short-budget profile, including all QMIX settings and all MAPPO settings except the large 15$\times$15-3p-4f task.
For MAPPO 15$\times$15-3p-4f, we use a $10$M-step medium-budget profile to provide a longer training horizon for the harder coordination setting.
Within each learner--environment setting, all compared methods use the same final training horizon.
Candidate-search methods share the same initial-search budget, and checkpoint-wise validation branches are treated as validation probes rather than independent final training trajectories.

\paragraph{Initial reward-shaping search.}
For both budget profiles, the initial reward-shaping search uses two rounds with three candidates per round.
Each candidate is trained for $850$K steps, and the selected initial endpoint is taken around $800$K steps.
This stage is used by candidate-search methods to identify an early shaping configuration.
Fixed-RS does not perform candidate search; it keeps a predefined shaping configuration fixed throughout training.
Sparse training uses only the original sparse environment reward.

\paragraph{Checkpoint-wise validation.}
For \textsc{LLM-ALSO}, checkpoint-wise validation is performed at two fixed intervention checkpoints.
At each checkpoint, the workflow evaluates two validation rounds with three update candidates per round.
A no-change control is automatically included at the checkpoint and does not count toward the update-candidate budget.
Thus, each checkpoint evaluates seven short-horizon branches in total: the no-change branch, three first-round update branches, and three second-round update branches.
This gives six effective update candidates per checkpoint and fourteen short-horizon validation branches across the two checkpoints.
Winner-branch promotion is enabled: the branch selected by the validation gate becomes the next mainline segment.

\begin{table}[t]
\centering
\small
\setlength{\tabcolsep}{4pt}
\begin{tabular}{lcc}
\toprule
Setting & 2.05M profile & 10M profile \\
\midrule
Final horizon & $2.05$M & $10$M \\
Sparse diagnostic run & $2.05$M & $10$M \\
Initial-search rounds & 2 & 2 \\
Candidates / round & 3 & 3 \\
Budget / candidate & $850$K & $850$K \\
Dense reference target & $2.05$M & $10$M \\
Adaptive mainline target & $2.05$M & $10$M \\
Validation checkpoints & $1.0$M, $1.5$M & $1.5$M, $2.5$M \\
Branch budget / branch & $300$K & $500$K \\
Rounds / checkpoint & 2 & 2 \\
Update candidates / round & 3 & 3 \\
Branches / checkpoint & 7 & 7 \\
Effective updates / checkpoint & 6 & 6 \\
Total branches over C1+C2 & 14 & 14 \\
Winner promotion & enabled & enabled \\
\bottomrule
\end{tabular}
\caption{
Workflow budgets for the short- and medium-budget adaptive training profiles.
The $2.05$M profile is used for all QMIX settings and MAPPO settings except 15$\times$15-3p-4f; the $10$M profile is used for MAPPO 15$\times$15-3p-4f.
}
\label{tab:workflow_budgets}
\end{table}

\paragraph{Baseline workflow accounting.}
Sparse training uses only the original sparse environment reward for the full final horizon.
Fixed-RS uses the same structured PBRS reward family as \textsc{LLM-ALSO}, but does not use LLM diagnosis or proposal, behavior evidence, branch validation, or checkpoint-wise adaptation.
It selects a fixed configuration from three pre-registered non-LLM static directions---balanced progress, early discovery, and collection readiness---and keeps the selected configuration unchanged throughout the full training horizon.
Single-LLM-RG uses one LLM to propose and refine structured PBRS candidates under the same initial-search budget as \textsc{LLM-ALSO}, selects one configuration using rule-based sparse-evaluation metrics, and then trains from scratch with this configuration fixed.
Thus, Fixed-RS controls for static structured reward shaping, Single-LLM-RG controls for single-LLM initial reward generation, and \textsc{LLM-ALSO} tests whether checkpoint-wise validation and adaptive replacement provide additional benefit.

\paragraph{Selected trajectory construction.}
For Sparse, Fixed-RS, and Single-LLM-RG, the reported trajectory is the corresponding final training run under the original sparse evaluation metric.
For \textsc{LLM-ALSO}, the reported selected trajectory follows the promoted mainline path.
It starts from the selected initial endpoint and then continues through any branch endpoints promoted by checkpoint-wise validation.
Unselected update branches are used only for validation and are not counted as final training trajectories.
This construction ensures that the reported average sparse return reflects the actual selected mainline rather than aggregating all validation probes.

\section{Reward-Shaping Configuration Details}
\label{app:pbrs_config_details}

\paragraph{Structured PBRS configuration space.}
All dense-signal methods use the same structured PBRS configuration space for LBF.
A reward-shaping configuration is represented by
\[
\theta=(\beta,m,\mathbf{w}),
\]
where $\beta$ is the shaping strength, $m$ is a high-level shaping mode, and $\mathbf{w}$ is a normalized weight vector over behavior-grounded potential components.
The runtime specification contains the implementation identifier, shaping mode, active component set, and per-component weights.
This design restricts LLM-generated candidates to an interpretable reward-shaping family rather than allowing arbitrary dense reward code.

\begin{table}[H]
\centering
\scriptsize
\setlength{\tabcolsep}{3pt}
\renewcommand{\arraystretch}{1.05}
\begin{tabular}{@{}p{0.13\columnwidth}p{0.28\columnwidth}p{0.51\columnwidth}@{}}
\toprule
Short name & Component & Main role \\
\midrule
col &
collection progress &
Progress toward successful food collection. \\
app &
approach behavior &
Movement toward feasible food targets. \\
cov &
coverage and discovery &
Exploration, map coverage, and food discovery. \\
ready &
joint-collection readiness &
Satisfaction of cooperative collection preconditions. \\
alloc &
target allocation &
Balanced assignment of agents to available targets. \\
stab &
late-stage stability &
Reduction of unstable or oscillatory late-stage behavior. \\
\bottomrule
\end{tabular}
\caption{
LBF structured PBRS potential components.
Candidates assign normalized weights to these components, and the active component set records the emphasized subset.
}
\label{tab:pbrs_components}
\end{table}

\paragraph{Configuration fields.}
A structured PBRS candidate contains two types of fields.
The runtime fields determine the actual shaping signal.
The implementation identifier specifies the structured PBRS implementation used in the LBF experiments; the mode specifies the high-level shaping direction; $\beta$ controls the global shaping strength; the active component set records which PBRS components are emphasized; and the weight vector gives the nonnegative per-component weights used to construct the potential.
These runtime fields are the authoritative reward-shaping specification used during training.

Other fields are used only as metadata for traceability and interpretation, including the candidate type, evidence keys, expected behavioral effect, and validation risk notes.
These metadata fields help audit LLM proposals, but they do not directly define the shaping potential.

\paragraph{Weight normalization and active components.}
The weight vector $\mathbf{w}$ is defined over the six LBF structured PBRS components listed in Table~\ref{tab:pbrs_components}.
Weights are nonnegative and normalized before use, so the active shaping potential remains within the structured component family.
The active component set records which components are intended to be emphasized in the candidate.
Components with zero or near-zero weight are not emphasized by the potential, even if the full weight vector contains entries for all six components for schema consistency.

\paragraph{Stage-dependent shaping modes.}
The shaping mode $m$ constrains a candidate toward a stage-appropriate reward-shaping direction.
For early sparse-reward learning, modes such as early-discovery or coverage-recovery emphasize coverage and approach behavior, encouraging exploration and target discovery.
Modes such as collection-readiness emphasize readiness and collection-progress terms, helping agents move from discovery to successful joint collection.
Later-stage modes such as late-stability reduce unnecessary exploration emphasis and focus on collection progress, allocation, readiness, or stability.
The mode is not an additional reward term; it is a high-level constraint that guides how component weights are selected.

\paragraph{Representative configuration behavior.}
A typical early-stage configuration assigns higher weights to coverage and approach components, with readiness as a secondary signal, so that agents are encouraged to discover feasible targets and move toward joint collection.
Later-stage configurations may reduce exploration-oriented weights and emphasize collection progress, allocation, readiness, or stability, depending on the diagnosed training stage.
Thus, candidates share the same component family but differ in shaping strength, mode, active component set, and normalized weights.

\paragraph{Runtime use and evaluation.}
Structured PBRS configurations are instantiated during training through $\beta$, the shaping mode, active components, and normalized component weights.
The shaping signal is disabled during evaluation, and all reported metrics use the original sparse environment reward.
Thus, the compared dense-signal methods differ in how they choose or update the training-time shaping configuration, while sharing the same sparse evaluation objective.

\clearpage
\twocolumn[
\section{Additional Analysis of Stage-Aware Adaptation}
\label{app:stage_adaptation_analysis}

\vspace{0.3em}
\begin{center}
\small
\setlength{\tabcolsep}{5pt}
\renewcommand{\arraystretch}{1.05}
\begin{tabular}{@{}lllrrll@{}}
\toprule
Checkpoint & Round & Branch & Score & Margin & Decision & Update promoted \\
\midrule
C1 & -- & NC    & 0.216 & 0.000  & local control & No \\
C1 & 1  & R1-u1 & 0.275 & 0.058  & evaluated & No \\
C1 & 1  & R1-u2 & 0.383 & 0.167  & selected & Yes \\
C1 & 1  & R1-u3 & 0.254 & 0.038  & evaluated & No \\
\midrule
C2 & -- & NC    & 0.765 & 0.000  & retained & No \\
C2 & 1  & R1-u1 & 0.627 & -0.138 & evaluated & No \\
C2 & 1  & R1-u2 & 0.756 & -0.009 & evaluated & No \\
C2 & 1  & R1-u3 & 0.767 & 0.002  & near tie; rejected & No \\
C2 & 2  & R2-u1 & 0.686 & -0.079 & evaluated & No \\
C2 & 2  & R2-u2 & 0.573 & -0.192 & evaluated & No \\
C2 & 2  & R2-u3 & 0.570 & -0.195 & evaluated & No \\
\bottomrule
\end{tabular}

\vspace{0.35em}
\captionof{table}{
Branch-level checkpoint validation scores for the representative QMIX 10$\times$10-2p-1f run.
NC denotes the mandatory no-change control, and margins are computed relative to the no-change score at each checkpoint.
}
\label{tab:checkpoint_validation_details}
\end{center}
\vspace{0.5em}
]

This appendix provides branch-level scoring details for the representative checkpoint-wise validation example analyzed in Section~\ref{sec:staged_analysis}.
The example corresponds to a QMIX run on LBF 10$\times$10-2p-1f.
The selected adaptive mainline is constructed from the selected initial-search endpoint and the subsequent checkpoint-wise validation decisions.
Along this trajectory, the workflow adopts an update branch at C1 and retains the no-change control at C2.

At each checkpoint, the current reward-shaping configuration is evaluated as a mandatory no-change control.
Update candidates are evaluated through short-horizon branches of 300K environment steps.
The corresponding branch trajectories and validation margins are visualized in Figures~\ref{fig:branch_validation_curves} and~\ref{fig:checkpoint_validation}; this appendix reports the score-level details used to produce those decisions.
For C1, the validation comparison includes the no-change control and the first-round update branches.
For C2, it includes all validation branches, with the no-change control and the near-tie update candidate being most relevant to the final decision.

We report branch validation using a stability-aware score-margin metric.
For each branch, the validation score summarizes sparse-evaluation performance over the short branch horizon rather than relying on a single peak value.
Specifically, the score combines the last-$k$ mean, short-horizon AUC, final return, and best return, while penalizing unstable or spiky behavior:
\begin{equation}
\label{eq:branch_score}
\begin{aligned}
S_j(b) ={}&
0.35\,\mathrm{LastK}_j(b)
+ 0.35\,\mathrm{AUC}_j(b) \\
&+ 0.20\,\mathrm{Final}_j(b)
+ 0.10\,\mathrm{Best}_j(b) \\
&- 0.10\bigl(
\mathrm{SpikeGap}_j(b)
+ \mathrm{StdLastK}_j(b)
\bigr).
\end{aligned}
\end{equation}
Here, $\mathrm{LastK}$ is the average over the last few evaluation points, $\mathrm{AUC}$ summarizes performance over the short validation horizon, $\mathrm{Final}$ is the last evaluation value, and $\mathrm{Best}$ is the maximum value observed during the branch.
The spike gap is defined as $\mathrm{Best}-\mathrm{LastK}$, penalizing candidates that only achieve a transient peak.

For each checkpoint $C_j$, the margin of branch $b$ is then computed relative to the local no-change control:
\[
\Delta_j(b) = S_j(b) - S_j(\text{no-change}),
\]
where $S_j(\text{no-change})$ is the score of the no-change branch at the same checkpoint.
A positive margin indicates improvement over the control, a negative margin indicates underperformance, and values near zero indicate a near tie.
The final replacement decision uses this score comparison together with risk-aware validation: stability, reference, and first-replacement risks are treated as penalties or warnings rather than automatic vetoes, while severe invalidity or near ties are still conservatively rejected.
Thus, an update is promoted only when its validation evidence clearly exceeds the no-change baseline after stability checks.

\paragraph{C1 decision.}
C1 is the post-initialization checkpoint at one million environment steps.
Among the evaluated C1 branches, R1-u2 achieves the largest margin over the no-change control, with a score margin of $0.167$.
The workflow therefore promotes this branch and continues from its endpoint.
This illustrates the proposal--adoption separation in \textsc{LLM-ALSO}: a candidate is adopted only when short-horizon validation supports improvement over the current configuration.

\paragraph{C2 decision.}
C2 is the late-stage checkpoint at 1.5 million environment steps.
Although R1-u3 is the strongest update candidate, its margin over the no-change control is only $0.002$, and the remaining updates are below the control.
The deterministic gate therefore retains no-change rather than forcing another reward-shaping update.

\paragraph{Interpretation.}
This example shows both outcomes of checkpoint-wise validation.
At C1, branch evidence is strong enough to promote an update into the mainline; at C2, the evidence is insufficient, so the current configuration is preserved.
Thus, checkpoint-wise validation acts as a selective replacement mechanism rather than an update-at-every-checkpoint heuristic.

\section{Learner Hyperparameters}
\label{app:learner_hyperparameters}

\begin{table}[H]
\centering
\scriptsize
\setlength{\tabcolsep}{3.5pt}
\renewcommand{\arraystretch}{1.00}
\begin{tabular}{@{}lcc@{}}
\toprule
Hyperparameter & Default QMIX & LBF-tuned QMIX \\
\midrule
Hidden dimension & 64 & 64 \\
Learning rate & $5\times10^{-4}$ & $3\times10^{-4}$ \\
Discount factor & 0.99 & 0.99 \\
Batch size & 32 & 32 \\
Replay buffer size & 5000 & 5000 \\
Mixer & QMIX & QMIX \\
Mixing embed dim. & 32 & 32 \\
Hypernet layers & 2 & 2 \\
Hypernet embed dim. & 64 & 64 \\
Double Q & True & True \\
Reward standardisation & True & True \\
Return standardisation & False & False \\
Network & FC / non-recurrent & GRU \\
Evaluation epsilon & 0.0 & 0.05 \\
Epsilon start & 1.0 & 1.0 \\
Epsilon finish & 0.05 & 0.05 \\
Epsilon anneal steps & 50K & 200K \\
Target update & 200 hard & 0.01 soft \\
Gradient clipping & 10 & 10 \\
Runner & episode & episode \\
MAC & basic\_mac & basic\_mac \\
Learner & q\_learner & q\_learner \\
\bottomrule
\end{tabular}
\caption{
QMIX hyperparameter presets.
Each preset is held fixed across all compared methods within the corresponding environment.
}
\label{tab:qmix_presets}
\end{table}

For each environment--algorithm pair, all compared methods use the same MARL learner hyperparameters.
Thus, performance differences are attributed to the training-time learning-signal design rather than to different learner configurations.

For QMIX, we use two explicitly recorded hyperparameter presets.
The default preset follows the QMIX configuration used in our codebase.
For the more coordination-demanding LBF settings, we use an LBF-tuned QMIX preset following the benchmark LBF QMIX setting~\citep{papoudakis2021benchmarking}.
Specifically, the default preset is used for 8$\times$8-2p-1f and 10$\times$10-2p-1f, while the LBF-tuned preset is used for 8$\times$8-2p-2f, 2s-8$\times$8-2p-2f, and 15$\times$15-3p-4f.
This preset choice affects only the QMIX learner and is held fixed across Sparse, Fixed-RS, Single-LLM-RG, and \textsc{LLM-ALSO} within each environment.
The 15$\times$15-3p-4f task is a large LBF task in the same environment family, so we apply the LBF-tuned preset rather than treating it as a separate task-specific tuned configuration.

For MAPPO, we use the default MAPPO learner configuration in our codebase for all LBF environments.
This configuration is shared by Sparse, Fixed-RS, Single-LLM-RG, and \textsc{LLM-ALSO}; only the training-time learning signal differs.
The MAPPO 15$\times$15-3p-4f setting uses the longer workflow budget described in Appendix~\ref{app:workflow_budgets}, while the remaining MAPPO settings use the short-budget profile.

\end{document}